\documentclass[preprint,aps]{revtex4}

\usepackage{graphicx}
\usepackage{dcolumn}
\usepackage{bm}
\usepackage{latexsym}
\usepackage{amsfonts}
\usepackage{amssymb}
\usepackage{amsmath}
\usepackage[usenames]{color}

\begin{document}
\preprint{KUNS-2425}

\title{On Cosmic No-hair in Bimetric Gravity and the Higuchi Bound}
\author{Yuki Sakakihara, Jiro Soda and Tomohiro Takahashi}
\affiliation{Department of Physics,  Kyoto University, Kyoto, 606-8501, Japan
}

\date{\today}

\begin{abstract}
We study the cosmic no-hair in the presence of spin-2 matter,
 i.e. in  bimetric gravity. 
We obtain stable de Sitter solutions with the cosmological constant in the physical sector 
and find an evidence that the cosmic no-hair is correct. 
In the presence of the other cosmological constant, 
there are two branches of de Sitter solutions.
Under anisotropic perturbations,
one of them is always stable and there is no violation of the cosmic no-hair at the linear level. 
The stability of the other branch depends on parameters and the cosmic no-hair can be violated in general. 
Remarkably, the bifurcation point of two branches exactly coincides with the Higuchi bound. 
It turns out that there exists a de Sitter solution for which the cosmic
 no-hair holds at the linear level and the effective mass for the anisotropic perturbations is above the Higuchi bound.
\end{abstract}

\maketitle

\section{Introduction}

It is well recognized that the large scale structure of the universe stems from primordial fluctuations generated quantum mechanically during inflation. 
Remarkably, the nature of primordial fluctuations is independent of initial conditions.  
This nice feature can be associated with the conjecture that the initial anisotropy and inhomogeneity rapidly disappear. This is called the cosmic no-hair conjecture.
The cosmic no-hair is proved in an ideal situation \cite{Wald:1983ky}.
Namely, a homogeneous expanding spacetime with a cosmological constant
rapidly approaches de Sitter spacetime, i.e., the initial anisotropy decays in a Hubble time, 
when we assume that matter satisfies the strong and dominant energy conditions.
In general, however, it is not clear whether the cosmic no-hair conjecture is correct or not. In fact, a counter example to this conjecture was found \cite{Watanabe:2009ct}. 
There, spin-1 gauge fields remain during inflation and the anisotropy does
not necessarily vanish. Moreover, it turned out that anti-symmetric
tensor fields can also generate the anisotropy
\cite{Watanabe:2010fh}. Hence, it is natural to explore the possibility
that a symmetric spin-2 tensor as matter causes the violation of the cosmic no-hair conjecture. Historically, a model of massive spin-2 matter has been proposed as that of meson \cite{Isham:1971gm}, which can be regarded as bimetric gravity consisting of the physical metric and the other spin-2 tensor field. 
In order to treat the spin-2 matter, therefore,
we need to construct a consistent ghost-free theory of bimetric gravity. 
Fortunately, this task has been accomplished recently \cite{deRham:2010ik,deRham:2010kj,Hassan:2011hr,Hassan:2011zd,Hassan:2011ea}.

Given a consistent model of spin-2 matter, we can study the cosmic no-hair conjecture.
There are some reasons that we expect the conjecture can be violated.
In the presence of spin-2 matter, 
it is inevitable that gravitons have the mass as the consequence of mixing between the physical metric and the other spin-2 tensor field. 
When we consider massive gravitons in an expanding spacetime,
the decay time scale of the anisotropy is determined by comparing Hubble scale with the effective mass of gravitons.
For example, by taking the couplings of the physical metric and the spin-2 matter small, the Hubble friction term might be dominant compared with the effective mass term in the equation of motion
then the decay time scale becomes much longer than Hubble time scale. 
Besides the above one, 
there may be the violation of the energy conditions in the presence of the spin-2 matter \cite{Volkov:2011an}.
Since the energy conditions are assumed in the proof of \cite{Wald:1983ky},
it is not apparent whether the cosmic no-hair holds or not in bimetric gravity.

In this paper, we consider a cosmological constant in bimetric gravity
as the limit of slow roll inflation. First, we concretely reveal the
property of de Sitter solutions in bimetric gravity. Then, we
investigate the fate of the anisotropy perturbatively. We stress that 
it is important to study the background geometry in detail because the effective mass of gravitons can depend on the background geometry.
Since, in known cases, the violation of the cosmic no-hair appears already at the linear level, we expect the linear analysis reflects the feature at the nonlinear level.

When we consider massive gravitons in de Sitter spacetime, 
we also need to care about the fact that the helicity-0 mode of massive gravitons
 becomes a ghost when
the effective mass is below the Higuchi bound \cite{Higuchi:1986py,deRham:2012kf,Fasiello:2012rw,Hassan:2012gz}.
Note that this ghost is different from a Boulware-Deser type ghost \cite{Boulware:1973my} which is already removed by construction.
Since there is no a priori reason to forbid the mass of gravitons violating the Higuchi bound, we also check if the effective mass satisfies the Higuchi bound. 

We organize the paper as follows.
In section \ref{sec:AB},
we present  ghost-free bimetric gravity
and derive basic equations needed for the analysis.
In section \ref{sec:L0},
we study the cosmic no-hair in bimetric gravity in the presence of the cosmological constant in the physical sector. 
We find that de Sitter solutions are stable and the small anisotropy rapidly decays. 
In section \ref{sec:LN0},
we introduce the other cosmological constant 
and investigate the stability of de Sitter solutions and  the fate of
the anisotropy.
We also study whether the Higuchi bound is satisfied or not.
The final section is devoted to the conclusion.  
In appendix A, we derive a set of equations used in the text.

\section{Bimetric gravity}\label{sec:AB}

In this section, we introduce bimetric gravity \cite{Hassan:2011zd,Hassan:2011ea} as a model of spin-2 matter and provide  basic formulae. 
Historically, after the pioneering work \cite{Isham:1971gm}, 
bimetric gravity has been studied from time to time \cite{Damour:2002wu,Damour:2002ws}. 
The model can be generalized to that of ghost-free multi-spin-2 matter \cite{Khosravi:2011zi,Nomura:2012xr}. 

Let us represent the physical metric and the other metric as $g_{\mu\nu}$ and $f_{\mu\nu}$, respectively.
Note that we regard the other metric $f_{\mu\nu}$ as the spin-2 matter.
We consider bimetric gravity with cosmological constants 
\begin{eqnarray}
	S&=&\frac{M_{g}^{2}}{2}\int d^{4}x \sqrt{-g}(R[g_{\mu\nu}]-2\Lambda_{g})
         +\frac{M_{f}^{2}}{2}\int d^{4}x \sqrt{-f}(R[f_{\mu\nu}]-2\Lambda_{f})\nonumber\\
        && +m^{2}M_{e}^{2}\int d^{4}x  \sqrt{-g} \sum_{n=1}^{3} 
              \alpha_n F_{n}[L^{\mu}_{\nu}] \ ,
\end{eqnarray}
where $M_g$ and $M_f$ are Planck constants of $g_{\mu\nu}$ and $f_{\mu\nu}$, 
and $R$ is the scalar curvature constructed from each metric.
The interaction terms of the metrics are defined as
\begin{equation*}
 F_{n}[X^{\mu}_{\nu}]
 =\frac{1}{n!}\sum_{\sigma\in S_{n}}\mathrm{sgn}(\sigma)
 X_{\mu_{1}}^{\mu_{\sigma(1)}}X_{\mu_{2}}^{\mu_{\sigma(2)}}\cdots
 X_{\mu_{n}}^{\mu_{\sigma(n)}} \ ,
\end{equation*}
\begin{equation*}
	 L_{\nu}^{\mu}= \delta_{\nu}^{\mu}-(\sqrt{g^{-1}f})_{\nu}^{\mu}
	  \ .
\end{equation*}
This combination of interaction terms gives no Boulware-Deser ghost \cite{Hassan:2011ea}.
Here, $m^2$ is a coupling constant of the metrics
and $\{\alpha_n\}_{n=1,2,3}$ are arbitrary constants.
We define the reduced Planck constant $M_e$ as 
\begin{equation*}
 	 \frac{1}{M_e^2}=\frac{1}{M_{g}^{2}}+\frac{1}{M_{f}^{2}} \ ,
\end{equation*}
where $M_e$ is chosen so that $m$ coincides with the Fierz-Pauli mass \cite{Fierz:1939ix} 
when we take the massive gravity limit. 
Note that we can regard $\Lambda_g$ as the potential energy of a scalar
field in the slow roll approximation coupled to the physical metric $g_{\mu\nu}$ as in general relativity.

In this paper, 
we consider the simplest case $\alpha_{2}=1$, $\alpha_{1}=\alpha_{3}=0$.
Then, the action is written as
\begin{eqnarray}
	S&=&\frac{M_{g}^{2}}{2}\int d^{4}x \sqrt{-g}(R[g_{\mu\nu}]-2\Lambda_{g})
	+\frac{M_{f}^{2}}{2}\int d^{4}x \sqrt{-f}(R[f_{\mu\nu}]-2\Lambda_{f})\nonumber\\
	&&+m^{2}M_{e}^{2}\int d^{4}x \sqrt{-g} F_{2}[L^{\mu}_{\nu}] \ ,
\end{eqnarray}
where
\begin{equation*}
	 F_{2}[L^{\mu}_{\nu}]=\frac{1}{2} ([L]^{2}-[L^{2}]) \ , \qquad
	 [L] = L_{\mu}^{\mu} \ ,\qquad
	 [L^{2}]=L_{\nu}^{\mu}L_{\mu}^{\nu} \ .
\end{equation*}

We now present basic equations and derive formulae which will be used in
the later analysis.

\subsection{de Sitter solutions in bimetric gravity}\label{hom-sol}

In this subsection, we consider homogeneous and isotropic solutions in bimetric gravity
\cite{Volkov:2011an,vonStrauss:2011mq,Comelli:2011zm,Volkov:2012cf,Berg:2012kn,Akrami:2012vf}.  
We derive equations of motion and show the solutions are  de Sitter spacetimes.

We take the homogeneous and isotropic metric ansatz for $g_{\mu\nu}$ and $f_{\mu\nu}$, 
\begin{equation}
	ds^{2}=-N^{2}(t)dt^{2}+e^{2\alpha(t)}[dx^{2}+dy^{2}+dz^{2}] \ ,
\end{equation}
and
\begin{equation}
	ds'^{2}=-M^{2}(t)dt^{2}+e^{2\beta(t)}[dx^{2}+dy^{2}+dz^{2}] \ ,
\end{equation}
respectively. 
$M$, $N$ are lapse functions and $\alpha$, $\beta$ describe the isotropic expansion of each metric.
Substituting the metric ansatz into the action, we obtain the Lagrangian
\begin{eqnarray}
 \mathcal{L}&=&
 M_{g}^{2}e^{3\alpha}\Bigl[-\frac{3\dot{\alpha}^{2}}{N}-N\Lambda_{g}\Bigr]
  +M_{f}^{2}e^{3\beta}\Bigl[-\frac{3\dot{\beta}^{2}}{M}-M\Lambda_{f}\Bigr]
\nonumber\\
&& +m^{2}M_{e}^{2}Ne^{3\alpha}\bigl[6-9\epsilon+3\epsilon^2+\gamma(-3+3\epsilon)\bigr] \ ,
\end{eqnarray}
where
\begin{eqnarray}
 \gamma = \frac{M}{N} \ , \qquad 
\epsilon=e^{\beta-\alpha} \ .
\end{eqnarray}
Taking the variation with respect to each variable, we obtain the equations of motion for $\alpha$ and $\beta$
\begin{equation}
  \Bigl(\frac{\alpha'}{N}\Bigr)' -\xi
   a_{g}(M-N\epsilon)\Bigl(\frac{3}{2}-\epsilon\Bigr)=0 \ ,
\label{eom_a}
\end{equation}
\begin{equation}
 \Bigl(\frac{\beta'}{M}\Bigr)'+\xi
  (1-a_{g})\epsilon^{-3}(M-N\epsilon)\Bigl(\frac{3}{2}-\epsilon\Bigr)=0
  \ ,
\label{eom_b}
\end{equation}
 and two constraints
\begin{equation}
  \Bigl(\frac{\alpha'}{N}\Bigr)^{2}=\lambda_g+\xi
   a_{g}(2-\epsilon)(\epsilon-1) \ ,\label{const1}
\end{equation}
\begin{equation}
  \Bigl(\frac{\beta'}{M}\Bigr)^{2}=\lambda_f+\xi
   (1-a_{g})\epsilon^{-3}(1-\epsilon) \ ,  \label{const2}
\end{equation}
where we normalized parameters and time with $M_e$ as follows:
\begin{equation}
 a_{g}=\frac{M_{e}^{2}}{M_{g}^{2}} \ ,\qquad
 \xi=\frac{m^{2}}{M_{e}^{2}} \ , \qquad
 \lambda_{g}=\frac{\Lambda_{g}}{3M_{e}^{2}} \ , \qquad
 \lambda_{f}=\frac{\Lambda_{f}}{3M_{e}^{2}} \ , \qquad
 ' = \frac{1}{M_e}\frac{\mathrm{d}}{\mathrm{d}t} \ .
\end{equation}
We notice that $a_g$ can take the value in the range $0<a_{g}<1$ from the definition of $M_e$.
The detailed derivation can be found in Appendix \ref{sec:DB}.

In bimetric gravity, the diagonal part of general coordinate invariance is preserved.
Hence, the two constraints contain a first class constraint 
and a second class constraint.
Thus, there exists a secondary constraint.
Now, from (\ref{eom_a}) and (\ref{const1}) (or (\ref{eom_b}) and (\ref{const2})),
we can deduce the equation
\begin{equation}
 \xi \Bigl(\frac{3}{2}-\epsilon\Bigr)\Bigl(\frac{\beta'e^{\beta}}{M}-\frac{\alpha'e^{\alpha}}{N}\Bigr)=0 \ .
\end{equation}
The first factor can be taken to be zero. However, this is a special solution and it is known that this leads to a pathology \cite{Gumrukcuoglu:2011zh,Comelli:2012db,
DeFelice:2012mx,Gumrukcuoglu:2012aa,Tasinato:2012ze}. 
Hence, we take the following branch
\begin{eqnarray}
 M =  \frac{\beta'}{\alpha'} N \epsilon \ .
\label{cons}
\end{eqnarray} 
This is nothing but the condition determining the Lagrange multiplier.
From (\ref{const1}), (\ref{const2}) and (\ref{cons}), we obtain the secondary
constraint
\begin{equation}
  g(\epsilon)=(\lambda_f+\xi a_{g})\epsilon^{3}-3\xi
 a_{g}\epsilon^{2}+[-\lambda_g+2\xi a_{g}-\xi(1-a_{g})]\epsilon+\xi
 (1-a_{g})=0 \ . \label{g}
\end{equation}
From the definition of $\epsilon$, $\epsilon$ should be positive and hence we 
should look for the positive roots of the algebraic equation $g(\epsilon)=0$.
Since $\xi$, $a_{g}$, $\lambda_g$ and $\lambda_f$ are constants, 
a positive root of $g(\epsilon)=0$ is also a constant which we represent $\epsilon_{0}$. 
Then, 
taking the derivative of the definition of $\epsilon$, 
we derive 
$\alpha'=\beta'$ and hence $M=N\epsilon_{0}$. 
Now, we take a gauge $N=1$ using the gauge degree of freedom. 
Then, we get $M= \epsilon_0= $ constant. 
From (\ref{eom_a}) and (\ref{eom_b}),
we can deduce $\alpha''=\beta''=0$ which can be solved as
$\alpha=H_{0} M_e t$, $\beta=H_{0} M_e t+\log{(\epsilon_{0})}$, where $H_0$ is Hubble scale
which is determined from the constraints as
\begin{eqnarray}
 H_{0}^{2}&=&\lambda_g+\xi a_{g}(2-\epsilon_{0})(\epsilon_{0}-1)\nonumber\\
&=&\lambda_f\epsilon_{0}^{2}+\xi(1-a_{g})\frac{1-\epsilon_{0}}{\epsilon_{0}}
 \ .\label{h02}
\end{eqnarray}
Thus, we obtained two de Sitter spacetimes with the relation $f_{\mu\nu}=\epsilon_0^2 \ g_{\mu\nu}$ provided that $\epsilon_0$ is a positive root of $g(\epsilon)=0$ 
and $H_0^2>0$ holds for $\epsilon_0$.

\subsection{Fate of the anisotropy}\label{linear_pert}

In this subsection, we consider the anisotropy perturbatively and examine how the anisotropy evolves.
We also derive the effective mass of the massive graviton.

We take the anisotropic metric ansatz
\begin{equation}
	ds^{2}=-N^{2}(t)dt^{2}
                +e^{2\alpha(t)}[e^{-4\sigma(t)}dx^{2}+e^{2\sigma(t)}(dy^{2}+dz^{2})]
                \ ,
\end{equation}
and
\begin{equation}
	ds'^{2}=-M^{2}(t)dt^{2}
                +e^{2\beta(t)}[e^{-4\lambda(t)}dx^{2}+e^{2\lambda(t)}(dy^{2}+dz^{2})]
                \ ,
\end{equation}
where $\sigma$ and $\lambda$ describes the anisotropic expansion of each metric.
Here we assume the anisotropy is small. 
Substituting the metric ansatz into the action and dropping the higher order terms, we can derive the quadratic Lagrangian
\begin{eqnarray}
 \delta^2 \mathcal{L}&=&
 M_{g}^{2}e^{3\alpha}\frac{3\dot{\sigma}^{2}}{N}
  +M_{f}^{2}e^{3\beta}\frac{3\dot{\lambda}^{2}}{M}
 +m^{2}M_{e}^{2}Ne^{3\alpha}\bigl[-9\epsilon+3\epsilon^2+3\gamma\epsilon]q^2
 \ ,
\end{eqnarray}
where we defined the new variable
\begin{equation}
 q = \lambda-\sigma \ .
\end{equation}
Note that $\sigma$ and $\lambda$ can be regarded as zero modes of
gravitons.
From the above action, we can deduce the equations for $\sigma$ as
\begin{equation}
 \sigma''+3H_{0}\sigma'-\xi a_{g}\epsilon_{0}(3-2\epsilon_{0})q=0 \ ,
\label{sigma}
\end{equation}
and  for $\lambda$ as 
\begin{equation}
 \lambda''+3H_{0}\lambda'+\xi (1-a_{g})\frac{1}{\epsilon_{0}}(3-2\epsilon_{0})q=0 \ .
\label{lambda}
\end{equation}
By taking the difference of (\ref{lambda}) and (\ref{sigma}), it is easy to obtain
\begin{equation}
 q''+3H_{0}q'+
 \xi\Bigl[a_{g}\epsilon_{0}+(1-a_{g})\frac{1}{\epsilon_{0}}\Bigr](3-2\epsilon_{0})q=0
 \ .
\label{eomq}
\end{equation}
From this equation, 
we can read off the effective mass of the massive graviton as
\begin{equation}
 m_{\rm eff}^{2}
 =\xi\Bigl[a_{g}\epsilon_{0}+(1-a_{g})\frac{1}{\epsilon_{0}}\Bigr](3-2\epsilon_{0}) \ .
\label{omega0}
\end{equation}
Since the effective mass is different from the bare mass $\xi$, 
it is non-trivial if the effective mass is less than Hubble scale even if the bare mass is so. 
By making the combination
$(\ref{sigma})\times 1/a_g + (\ref{lambda}) \times\epsilon_0^2 /(1-a_g)$, we have
\begin{equation}
 \biggl[e^{3H_{0}t}\Bigl(\frac{\sigma'}{a_{g}}+\epsilon_{0}^{2}\frac{\lambda'}{1-a_{g}}\Bigr)
\biggr]'=0 \ .
\label{massless}
\end{equation}
This leads to a conserved quantity
\begin{equation}
 E =  e^{3H_{0}t}\Bigl(\frac{\sigma'}{a_{g}}+\epsilon_{0}^{2}\frac{\lambda'}{1-a_{g}}\Bigr) \ 
\label{cq}
\end{equation}
which means the mode
\begin{eqnarray}
\frac{\sigma'}{a_{g}}+\epsilon_{0}^{2}\frac{\lambda'}{1-a_{g}}
\end{eqnarray}
corresponds to the massless graviton. The existence of the massless mode is
a reflection of the diagonal general coordinate invariance.
From the conservation law (\ref{cq}), we see that this mode vanishes exponentially fast.

If we substitute $q=e^{i\omega t}$ into eq. (\ref{eomq}), we obtain
\begin{equation}
 q=A\exp{i\omega_+ t}+B\exp{i\omega_- t} \ ,
\end{equation}
where
\begin{equation}
 \omega_{\pm}=i\frac{3H_0}{2}\pm \sqrt{m_{\rm eff}^{2}-\frac{9 H_0^{2}}{4}}
\end{equation}
and $A,B$ are integral constants.
If $m_{\rm eff}^2$ is negative, $q$ exponentially grows like
\begin{equation}
 q \sim B \exp{\Bigl(\sqrt{|m_{\rm
  eff}^2|+\frac{9H_0^2}{4}}-\frac{3H_0}{2}\Bigr) t} \ .
\end{equation}
Inversely, if $m_{\rm eff}^2$ is positive, $q$ exponentially decays.
When $m_{\rm eff}^{2}-\frac{9H_0^2}{4}>0$,
the decay time scale $\tau$ is $\tau=2/3H_0$.
On the other hand, if $m_{\rm eff}^{2}-\frac{9H_0^2}{4}<0$, 
the time scale is evaluated as
\begin{equation}
 \tau^{-1}=|\omega_-|=\frac{3H_0}{2}-
  \sqrt{\frac{H_0^{2}}{4}+(2H_0^2-m_{\rm eff}^2)} \ .
\end{equation}
Therefore, the decay time scale of the anisotropy $\tau$ is shorter than Hubble time scale $1/H_0$ for $m_{\rm eff}^2>2H_0^2$
and the opposite holds for $m_{\rm eff}^2<2H_0^2$.

\section{Decay of the anisotropy: cases $\lambda_f =0$}\label{sec:L0}

First, we consider the situation $\lambda_f=0$. 
The constant $\lambda_g$ can be regarded as the potential energy of a scalar field coupled to $g_{\mu\nu}$ in the slow roll approximation.
We prove that there exist a de Sitter solution for $\lambda_g>0$
and the solution is stable under the anisotropic perturbations.
We also see that the effective mass of the massive graviton is bounded from below $m_{\rm eff}^2 > 3H_0^2$.
This suggests that the anisotropy rapidly decays in a Hubble time.

When we take $\lambda_f=0$, (\ref{g}) and (\ref{h02}) become
\begin{equation}
  g(\epsilon)=\xi a_{g}\epsilon^{3}-3\xi
 a_{g}\epsilon^{2}+[-\lambda_g+2\xi a_{g}-\xi(1-a_{g})]\epsilon+\xi
 (1-a_{g})=0 \label{g-0}
\end{equation}
and
\begin{eqnarray}
 H_{0}^{2}&=&\lambda_g+\xi a_{g}(2-\epsilon_{0})(\epsilon_{0}-1) \nonumber\\
&=&\xi(1-a_{g})\frac{1-\epsilon_{0}}{\epsilon_{0}}\label{h02-0} \ .
\end{eqnarray}
From the second line of (\ref{h02-0}), we see that $\epsilon_0$ should be
less than 1 so that $H_0$ is a real number.
Then, from the first line of (\ref{h02-0}), $\lambda_g$ should have a
positive lower bound. 
We assume that $\lambda_g$ is positive in the following.
Then, we obtain 
\begin{equation*}
 g(0)=\xi(1-a_g)>0 \ ,\qquad 
 g(1)=-\lambda_g<0 \ , \qquad
 g(\epsilon)\rightarrow + \infty  \;\; \mathrm{as} \;\;  \epsilon\rightarrow +\infty
\ .
\end{equation*}
Thus, there are a positive root smaller than 1 
and a positive root larger than 1. 
The root larger than 1 does not satisfy the condition $H_0^2 >0$.
It turned out that 
there is a single positive root $\epsilon_0$ in the range $ 0<
\epsilon_0 < 1 $ where $H_0$ is a real number in the case $\lambda_g$ is
positive.

Next, we will see that the de Sitter solution derived above
is always stable under the anisotropic perturbations.
Apparently, massless modes rapidly decay in Hubble time scale.
Then the stability under the perturbations of $\sigma$ and $\lambda$ 
is determined by the sign of the mass term of the perturbation equation (\ref{eomq}) as mentioned in Sec. \ref{linear_pert}
or the sign of
\begin{equation*}
 m_{\rm eff}^{2}
=\xi\Bigl[a_{g}\epsilon_{0}+(1-a_{g})\frac{1}{\epsilon_{0}}\Bigr](3-2\epsilon_{0})
\ .
\end{equation*}
Since the de Sitter solution satisfies $0 < \epsilon_0 <1$, 
$m_{\rm eff}^2$ is positive. 
Therefore, the de Sitter solution is stable under the perturbations of $\sigma$ and $\lambda$.

Furthermore,
we can prove that $m_{\rm eff}^2$ is bounded from below $m_{\rm eff}^2 > 3H_0^2$.
To show this, let us define 
\begin{equation}
 h(\epsilon_{0})=\frac{m_{\rm eff}^{2}}{H_{0}^{2}}=\frac{[a_{g}\epsilon_{0}^{2}+(1-a_{g})](3-2\epsilon_{0})}{(1-a_{g})(1-\epsilon_{0})} \ ,
\end{equation}
where we used the second line of (\ref{h02-0}). 
It is straightforward to calculate the derivative of $h(\epsilon_0)$,
\begin{equation}
 \frac{\mathrm{d}}{\mathrm{d}\epsilon_{0}}h(\epsilon_{0})
=\frac{4a_{g}\epsilon_{0}(\epsilon_0-\frac{9}{8})^2+a_g \epsilon_0\frac{15}{16}+(1-a_g)}
{(1-a_{g})(1-\epsilon_{0})^{2}}  \ .
\end{equation}
Since this is manifestly positive in the range $0<\epsilon_{0}<1$, we
have the inequality $h(\epsilon_{0})>h(0)=3$, that is,
\begin{equation}
 m_{\rm eff}^{2}>3H_{0}^{2} \ .
\end{equation} 
The effective mass of the massive graviton is bounded by Hubble scale from below.
Using the analysis in Sec. \ref{linear_pert},
we can see that the anisotropy rapidly decays in a Hubble time.

\section{Decay of the anisotropy: cases $\lambda_f \neq 0$}\label{sec:LN0}

In this section, we construct de Sitter solutions with $\lambda_f\neq 0$.
Then, we check the perturbative stability of the de Sitter solutions.
Finally, we examine if the effective mass of the massive graviton can 
be smaller than Hubble scale.

\subsection{de Sitter solutions}

We study de Sitter solutions and give a classification of them.
What we should check is 
whether roots of $g(\epsilon)=0$ are positive and satisfy $H_0^2>0$.

\subsubsection{When are roots of $g(\epsilon)=0$ positive?}

Since the behavior of $g(\epsilon)$ is largely determined 
by the leading term,  $\lambda_f+\xi a_g$, we discuss the following three cases separately.

\begin{enumerate}

\item In the case $\lambda_f>-\xi a_{g}$,
the coefficient of the leading term in $g(\epsilon)$ is positive, which indicates
\begin{equation*}
 g(\epsilon)\rightarrow - \infty  \;\; \mathrm{as} \;\;  \epsilon\rightarrow -\infty \ , \qquad
 g(\epsilon)\rightarrow + \infty  \;\; \mathrm{as} \;\; \epsilon\rightarrow +\infty \ .
\end{equation*}
Combining the above with  $g(0)=\xi (1-a_{g})>0$, we see that there always exists a negative root. 
Since $g''(0)=-6\xi a_g<0$, the inflection point must exist in the positive side of $\epsilon$.  
Therefore, the number of positive solutions can be characterized by the discriminant of $g(\epsilon)=0$.
If the discriminant is zero, a multiple positive root exists. 
On the other hand, if the discriminant is positive, two positive roots exist.
The discriminant of $g(\epsilon)=0$ is given by
\begin{eqnarray}
 D&=& -27(1-a_g)^2\Bigl(\frac{\lambda_f}{\xi}+a_g\Bigr)^2 \nonumber\\
&& \quad +2\bar{c}[2\bar{c}^2+27a_g(1-a_g)]\Bigl(\frac{\lambda_f}{\xi}+a_g\Bigr)
+9a_g^2[\bar{c}^2+12a_g(1-a_g)]  \ ,
\end{eqnarray}
where we defined
\begin{equation}
 \bar{c}=\lambda_g/\xi-2a_{g}+(1-a_{g}) \ .
\end{equation}
The condition that the discriminant is non negative reads
\begin{equation}
 \lambda_{-}\leq\lambda_f\leq\lambda_{+} \ ,
\end{equation}
where we defined
\begin{equation}
 \frac{\lambda_{\pm}}{\xi}+ a_{g}
 =\frac{1}{27(1-a_{g})^{2}}
 \bigl\{\bar{c}[2\bar{c}^{2}+27a_{g}(1-a_{g})]
 \pm 2[\bar{c}^{2}+9a_{g}(1-a_{g})]^{\frac{3}{2}}\bigr\} \ .
\label{lpm}
\end{equation}
We can see $\lambda_{-}<-\xi a_{g}$ and $\lambda_{+}>-\xi a_{g}$
from (\ref{lpm}) taking into account the inequality
\begin{equation*}
 \big|2[\bar{c}^{2}+9a_{g}(1-a_{g})]^{\frac{3}{2}}\big|
-\big|\bar{c}[2\bar{c}^{2}+27a_{g}(1-a_{g})]\big|>0 \ .
\end{equation*}
Thus, for $\lambda_f=\lambda_+$, there exists a single multiple positive
      root of $g(\epsilon)=0$. 
Since we are considering the range $\lambda_f>-\xi a_g$,
there exist two positive roots for $-\xi a_g<\lambda_f<\lambda_{+}$.

\item In the case $\lambda_f=-\xi a_{g}$,
$g(\epsilon)$ becomes the quadratic function of $\epsilon$.
Since the coefficient of the leading term $-3\xi a_g$ is negative
and $g(0)=\xi (1-a_{g})>0$, there exists a single positive root.

\item In the case $\lambda_f<-\xi a_{g}$,
the coefficient of the leading term in $g(\epsilon)$ is negative, which leads to
\begin{equation*}
 g(\epsilon)\rightarrow + \infty  \;\; \mathrm{as} \;\;  \epsilon\rightarrow -\infty \ , \qquad
 g(\epsilon)\rightarrow - \infty  \;\; \mathrm{as} \;\;  \epsilon\rightarrow +\infty \ .
\end{equation*}
Because of the fact $g(0)=\xi (1-a_{g})>0$, there always exists a positive root. 
Since $g''(0)= -6\xi a_g < 0$,
the inflection point exists on the negative side of $\epsilon$ in this case.
Thus, other possible roots should be negative.
Namely, there exists a single positive root for $\lambda_f<-\xi a_g$.

\end{enumerate}

We found that two positive roots exist for $-\xi a_g<\lambda_f<\lambda_+$
and a single positive root exists for $\lambda_f\leq -\xi a_g$ and $\lambda_f=\lambda_+$.

Next, we check whether these roots satisfy the condition $H_0^2>0$.

\subsubsection{Is $H_{0}^{2}>0$ satisfied ?} \label{subs-h02}

Rewriting the first line of (\ref{h02}) as
\begin{eqnarray}
H_0^2 
&=& \lambda_g+\xi a_{g}(2-\epsilon_{0})(\epsilon_{0}-1)\notag \\
&=& \xi a_g\Bigl[-\Bigl(\epsilon_0-\frac{3}{2}\Bigr)^2
+\frac{\lambda_g}{\xi a_g}+\frac{1}{4}\Bigr] \ ,
\label{H2>0}
\end{eqnarray}
we see that $\lambda_g>-\xi a_{g}/4$ is at least needed for $H_0^2>0$.
Therefore we assume $\lambda_g>-\xi a_{g}/4$ below.
Then, we can factorize (\ref{H2>0}) as
\begin{equation}
 H_0^2 = -\xi a_g (\epsilon_0-\epsilon_{p})(\epsilon_0-\epsilon_{m}) \ ,
\end{equation}
where we defined
\begin{equation}
  \epsilon_{p}=\frac{3}{2}+\sqrt{\frac{\lambda_g}{\xi a_{g}}+\frac{1}{4}} \ ,\quad
 \epsilon_{m}=\frac{3}{2}-\sqrt{\frac{\lambda_g}{\xi a_{g}}+\frac{1}{4}} \ .\label{ecpm}
\end{equation}
Note that $\epsilon_{p}$ and $\epsilon_{m}$ do not depend on $\lambda_f$.
Thus, in order to have $H_0^2>0$, we have to seek positive roots of $g(\epsilon)=0$ in the range
\begin{equation}
   \epsilon_{m}<\epsilon_{0} < \epsilon_{p} \ .
\end{equation}

As we discussed in the previous subsection,
$\lambda_f\leq \lambda_+$ is needed for the existence of positive roots.
We first consider the case $\lambda_f=\lambda_+$ for which there exists a single positive root. 
In this case, we have to solve 
$g(\epsilon_*) = g' (\epsilon_*) = 0$ which give rise to the equation
\begin{eqnarray}
  a_g \epsilon_*^2 + \frac{2}{3} \bar{c} \epsilon_* -(1-a_g) =0 \ . 
\label{multiple}
\end{eqnarray}
The positive root of this equation is given by
\begin{equation}
 \epsilon_*=\frac{-\bar{c}+\sqrt{\bar{c}^2+9a_g(1-a_g)}}{3a_g} \ .
\label{e*}
\end{equation}
Thus, we see
\begin{equation}
 H_0^2(\epsilon_*)
 =\frac{2\xi (\frac{\lambda_g}{\xi}+\frac{a_g}{4})\sqrt{\bar{c}^2+9a_g(1-a_g)}}
 {(\frac{\lambda_g}{\xi}+\frac{a_g}{4})+\frac{9}{4}a_g+(1-a_g)+\sqrt{\bar{c}^2+9a_g(1-a_g)}}>0
 \ .
\end{equation}
Therefore, the inequality $\epsilon_{m}<\epsilon_*<\epsilon_{p}$ must hold.

As we decrease $\lambda_f$ with fixing $\lambda_g , a_g , \xi$, 
the discriminant of $g(\epsilon)=0$ becomes positive.
Thus, there will be two positive roots
until $\lambda_f$ reaches $-\xi a_g$.
We shall call smaller one inner root, $\epsilon_{\mathrm{in}}$,
 and the other one outer root, $\epsilon_{\mathrm{out}}$.
We note that 
$\epsilon_{\mathrm{in}}$ is always smaller than $\epsilon_*$
and $\epsilon_{\mathrm{out}}$ is always larger than $\epsilon_*$
because
\begin{equation*}
 g(0)=\xi (1-a_g)>0, \qquad
 g(\epsilon_*)=\epsilon_*^3(\lambda_f-\lambda_+)<0, \qquad
  g(\epsilon)\rightarrow + \infty  \;\; \mathrm{as} \;\; \epsilon\rightarrow +\infty.
\end{equation*}
We can regard $\lambda_f \leq-\xi a_g$ case as the inner root
because the inner root is continuously connected to the positive root for $\lambda_f <-\xi a_g$
when $\lambda_f$ crosses $-\xi a_g$ below.

We shall evaluate the first derivative of $\epsilon_0$ with respect to $\lambda_f$
since we want to know the behavior of the roots 
when we decrease $\lambda_f$.
Differentiating $g(\lambda_f,\epsilon_{0}(\lambda_f))=0$ with respect to $\lambda_f$ 
\begin{eqnarray}
  \frac{\mathrm{d} g(\lambda_f,x(\lambda_f))}{\mathrm{d}\lambda_f}\bigg|_{x=\epsilon_{0}}=0 \ ,
\end{eqnarray}
we obtain
\begin{eqnarray}
    \frac{\mathrm{d}\epsilon_{0}}{\mathrm{d}\lambda_f}
   =-\frac{\epsilon_{0}^{3}}{\frac{\mathrm{d}g(x)}{\mathrm{d}x}\big|_{x=\epsilon_{0}}} \ .
\label{dedlf}
\end{eqnarray}

\begin{figure}[!h]
\centering \includegraphics[height=3.5in]{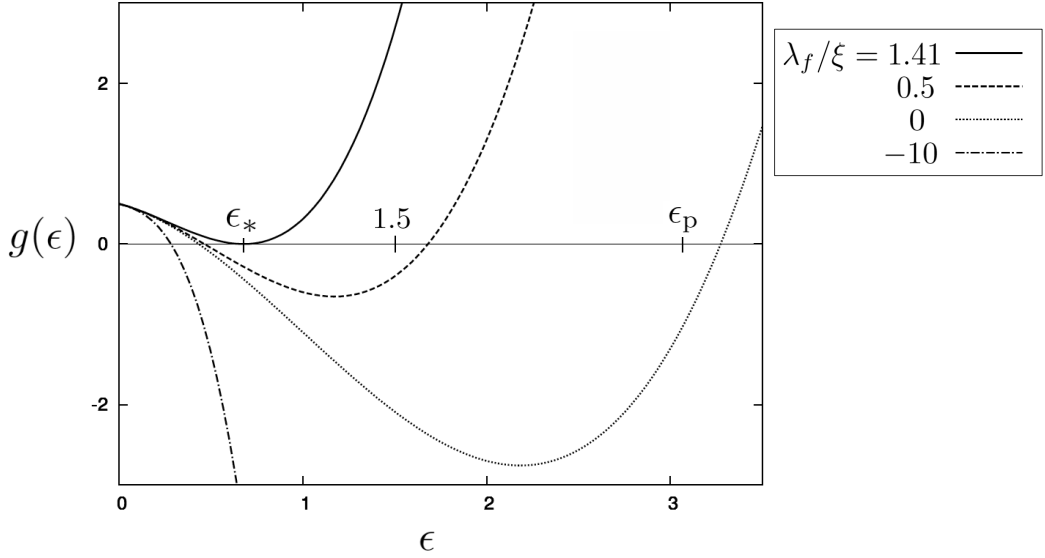}
\caption{We plotted $g(\epsilon)$ for $\lambda_g\geq 2\xi a_g$.
We set $a_g=0.5$, $\lambda_g/\xi=1.1$. 
Then $\lambda_+/\xi \simeq 1.41$.
As $\lambda_f$ decreases,
the outer root increases and the inner root decreases.
When $\lambda_f$ reaches $\lambda_{p}$, the outer root crosses $\epsilon_{p}$ above 
and $H_0^2(\epsilon_{\mathrm{out}})$ becomes negative.
But the inner root always satisfies $H_0^2(\epsilon_{\mathrm{in}})>0$ 
since $\epsilon_{m}$ is non positive.}
\label{fig.1}
\end{figure}

First, we discuss the outer root.
Since $g(\epsilon_*)<0$ and $g(\epsilon)\rightarrow +\infty$ as $\epsilon\rightarrow +\infty$,
the outer root always satisfies
\begin{equation}
  \frac{\mathrm{d}g(x)}{\mathrm{d}x}\bigg|_{x=\epsilon_{\mathrm{out}}}>0
   \ .
\end{equation}
Then, from (\ref{dedlf}), we can see
\begin{equation}
 \frac{\mathrm{d}\epsilon_{\mathrm{out}}}{\mathrm{d}\lambda_f}<0 \ .
\end{equation}
Therefore, $\epsilon_{\mathrm{out}}$ starts from $\epsilon_*$ at $\lambda_f=\lambda_+$
and 
$\epsilon_{\mathrm{out}}$ monotonically increases as $\lambda_f$ decreases.
We can expect that  $\epsilon_{\mathrm{out}}$ sometime reaches $\epsilon_{p}$.
Indeed, $\epsilon_{\mathrm{out}}$ reaches $\epsilon_{p}$ 
when $\lambda_f$ becomes small as 
\begin{equation}
 \lambda_{p}=\xi(1-a_{g})\frac{\epsilon_{p}-1}{\epsilon_{p}^{3}}>0 \ ,
\end{equation}
where we used the fact $H_0^2 =0 $ at $\epsilon_p$.
Therefore, $\epsilon_{\mathrm{out}}$ exists in the range $(\epsilon_{m},\epsilon_{p})$ 
if and only if $\lambda_f>\lambda_{p}$.
We mention that $\lambda_{p}\rightarrow +0$ when $\lambda_g\rightarrow +\infty$
since $\epsilon_{p}\rightarrow +\infty$ (see (\ref{ecpm})).

\begin{figure}[!h]
\centering \includegraphics[height=3.5in]{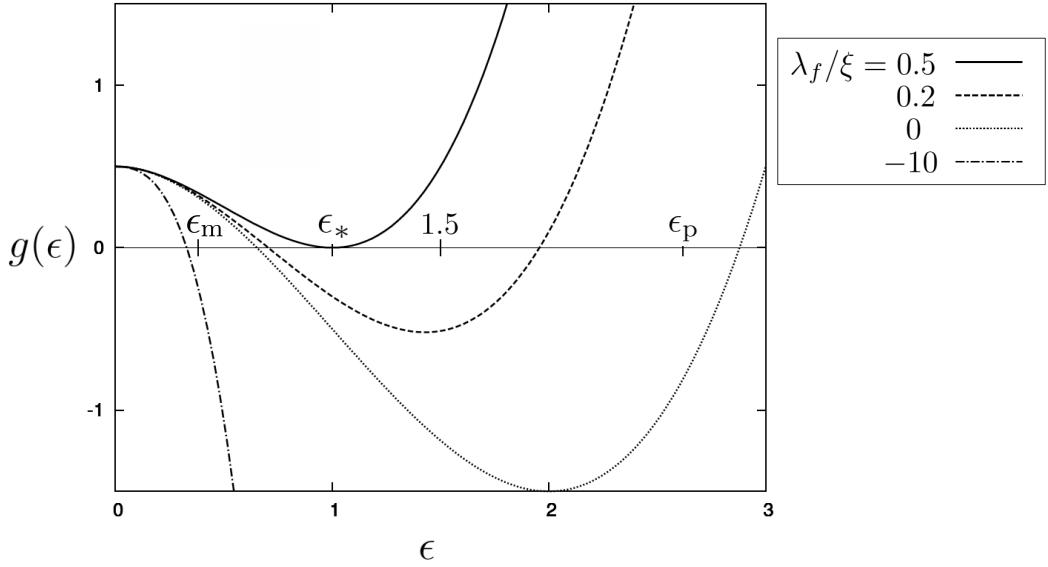}
\caption{We plotted $g(\epsilon)$ for $-\frac{1}{4}\xi a_g<\lambda_g<2\xi a_g$.
We set $a_g=0.5$, $\lambda_g/\xi=0.5$. 
Then $\lambda_+/\xi =0.5$. 
As $\lambda_f$ decreases, 
the outer root increases and the inner root decreases.
When $\lambda_f$ reaches $\lambda_{p}$, the outer root crosses $\epsilon_{p}$ above 
and $H_0^2(\epsilon_{\mathrm{out}})$ becomes negative.
When $\lambda_f$ reaches $\lambda_{m}$, the inner root crosses $\epsilon_{m}$ below 
and $H_0^2(\epsilon_{\mathrm{in}})$ becomes negative.
}
\label{fig.2}
\end{figure}

Next, we discuss the inner root.
In turn, since $g(0)=\xi (1-a_g)>0$ and $g(\epsilon_*)<0$, the inner root always satisfies
\begin{equation}
  \frac{\mathrm{d}g(x)}{\mathrm{d}x}\bigg|_{x=\epsilon_{\mathrm{in}}} <0
   \ .
\end{equation}
Then from (\ref{dedlf}), we can see
\begin{equation} 
 \frac{\mathrm{d}\epsilon_{\mathrm{in}}}{\mathrm{d}\lambda_f}>0 \ .
\end{equation}
Therefore, $\epsilon_{\mathrm{in}}$ starts from $\epsilon_*$ at $\lambda_f=\lambda_+$
and monotonically decreases as $\lambda_f$ decreases.
Note that
$\epsilon_{\mathrm{in}}\rightarrow\bigl(\frac{\xi(1- a_g)}{|\lambda_f|}\bigr)^{1/3}\rightarrow+0$
as $\lambda_f \rightarrow-\infty$.
We can expect that  $\epsilon_{\mathrm{in}}$ sometime reaches $\epsilon_{m}$.
To see this,
we need to notice that
\begin{equation}
 \epsilon_{m}=\frac{3}{2}-\sqrt{\frac{\lambda_g}{\xi a_g}+\frac{1}{4}}
=\frac{3}{2}-\sqrt{\frac{\lambda_g-2\xi a_g}{\xi a_g}+\frac{9}{4}}
\label{ecm}
\end{equation}
changes the sign at $\lambda_g=2\xi a_g$. Hence, we can consider the following two cases.
\begin{enumerate}
 \item In the case $\lambda_g\geq 2\xi a_g$,
$\epsilon_{m}$ is non positive.
Then $\epsilon_{\mathrm{in}}$ cannot reach $\epsilon_{m}$ when we decrease $\lambda_f$.
Therefore, $\epsilon_{\mathrm{in}}$ always exists in the range $(\epsilon_{m},\epsilon_{p})$
and satisfies $H^2_0(\epsilon_{\mathrm{in}})>0$ (see Fig. \ref{fig.1}).

 \item In the case $-\frac{1}{4}\xi a_g<\lambda_g<2\xi a_g$,
$\epsilon_{m}$ is positive.
Then $\epsilon_{\mathrm{in}}$ can reach $\epsilon_{m}$ when we decrease $\lambda_f$.
Indeed, $\epsilon_{\mathrm{in}}$ reaches $\epsilon_{m}$ 
when $\lambda_f$ becomes small as
\begin{equation}
 \lambda_{m}=\xi(1-a_{g})\frac{\epsilon_{m}-1}{\epsilon_{m}^{3}} \ , 
\end{equation}
where we used the fact $H_0^2 =0 $ at $\epsilon_m$.
Therefore, $\epsilon_{\mathrm{in}}$ exists in the range $(\epsilon_{m},\epsilon_{p})$ 
and satisfies $H_0^2>0$ if and only if $\lambda_f>\lambda_{m}$.
We mention that $\lambda_{m}\rightarrow -\infty$ when $\lambda_g\rightarrow 2\xi a_g -0$
because $\epsilon_{m}\rightarrow +0$ (see (\ref{ecm})).
In Fig. \ref{fig.2}, we illustrate these features.

\end{enumerate}

We note that $\lambda_{p}>\lambda_{m}$ when $-\frac{1}{4}\xi a_g<\lambda_g<2\xi a_g$.
We can see this from the definitions of $\lambda_{p}$ and $\lambda_{m}$ as
\begin{equation}
 \lambda_{p}-\lambda_{m}
=\xi(1-a_g)\frac{8\bigl(\frac{\lambda_g}{\xi a_{g}}+\frac{1}{4}\bigr)^{\frac{3}{2}}}
{\bigl(2-\frac{\lambda_g}{\xi a_{g}}\bigr)^{3}}>0 \ .
\end{equation}

\begin{figure}[!h]
\centering \includegraphics[height=3.5in]{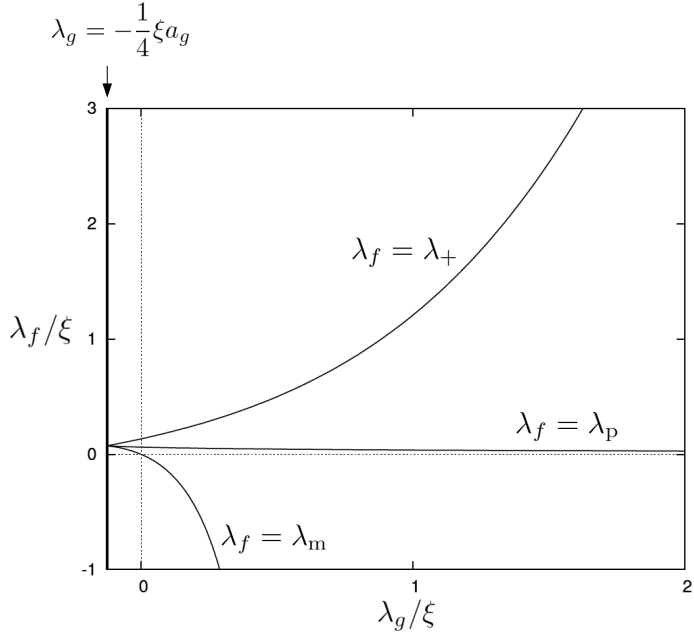}
\caption{We depicted the region de Sitter solutions exist.
We chose $a_g=0.5$.
The multiple solution exists on $\lambda_f=\lambda_+$ curve.
The outer root exists in the region 
below $\lambda_+$ and above $\lambda_{p}$.
The inner root exists in the region 
below $\lambda_+$ and above $\lambda_{m}$.
$\lambda_{p}\rightarrow +0$ as $\lambda_g \rightarrow +\infty$
and
$\lambda_{m}\rightarrow -\infty$ as $\lambda_g \rightarrow 2\xi a_g-0$
as we mentioned in the text.
The triple point is given by ($\lambda_g$, $\lambda_f$)
=($-\frac{1}{4}\xi a_g$, $\frac{4}{27}\xi (1-a_g)$) and there $H_0^2=0$.}
\label{fig.3}
\end{figure}

We summarize the results derived in this subsection in Table
\ref{table.1}, Table \ref{table.2} and Fig. \ref{fig.3}.

\begin{table}[!h]
\begin{minipage}[t]{.45\textwidth}
\caption{For $\lambda_g\geq 2\xi a_g$}
\begin{center}
\begin{tabular}{|c|c|c|}
 \hline
 &inner  &outer \\ \hline
 $\lambda_{+}<\lambda_f$ & $\times$ & $\times$ \\ \hline
 $\lambda_f=\lambda_{+}$ & \multicolumn{2}{c|}{$\circ$} \\ \hline 
 $\lambda_{p}<\lambda_f<\lambda_{+}$ & $\circ$ & $\circ$ \\ \hline
 $\lambda_f\leq\lambda_{p}$ & $\circ$ & $\times$ \\ 
 \hline
\end{tabular}
\label{table.1}
\end{center}		  
\end{minipage}
\hfill
\begin{minipage}[t]{.45\textwidth}
\caption{For $-\frac{1}{4}\xi a_g<\lambda_g<2\xi a_g$}
\begin{center}
\begin{tabular}{|c|c|c|}
 \hline
 &inner  &outer \\ \hline
 $\lambda_{+}<\lambda_f$ & $\times$ & $\times$ \\ \hline
 $\lambda_f=\lambda_{+}$ & \multicolumn{2}{c|}{$\circ$} \\ \hline 
 $\lambda_{p}<\lambda_f<\lambda_{+}$ & $\circ$ & $\circ$ \\ \hline
 $\lambda_{m}<\lambda_f\leq\lambda_{p}$ & $\circ$ & $\times$ \\ \hline
 $\lambda_f\leq\lambda_{m}$ & $\times$ & $\times$ \\ 
 \hline
\end{tabular}
\label{table.2}
\end{center}
\end{minipage}
\end{table}
In the tables, ``$\circ$'' means
there exists a positive root of $g(\epsilon)=0$ which satisfies $H_0^2>0$,
i.e., a de Sitter solution exists.
And, ``$\times$'' means 
there exists no positive root or
there exists a positive root for $g(\epsilon)=0$ but $H_0^2\leq 0$,
i.e., no de Sitter solution exists.
For $\lambda_g\leq-\frac{1}{4}\xi a_g$, there is no root satisfying $H_0^2>0$.
Surprisingly, we have an upper bound for $\lambda_f$ and there exist de Sitter solutions 
even for arbitrary large negative $\lambda_f$ in the case $\lambda_g \geq 2\xi a_g$. 
We note that $\lambda_g-2\xi a_g$ can be interpreted as an effective cosmological constant 
if we see the explicit constant term in the first line of (\ref{h02}).
It is remarkable that there also exists a de Sitter solution for the case
effective cosmological constant is zero.

\subsection{Stability of de Sitter solutions}

In this subsection, 
we  examine the stability of de Sitter solutions.
In Sec. \ref{linear_pert},
We saw that the sign of $m_{\rm eff}^2$ determines the stability of de Sitter solutions,
i.e., solutions are stable if $m_{\rm eff}^2$ is positive
and unstable if $m_{\rm eff}^2$ is negative.
Recalling the formula
\begin{equation*}
 m_{\rm eff}^{2}
=\xi\Bigl[a_{g}\epsilon_{0}+(1-a_{g})\frac{1}{\epsilon_{0}}\Bigr](3-2\epsilon_{0})
\ ,
\end{equation*}
we can see that
$m_{\rm eff}^2$ is positive when $\epsilon_0<\frac{3}{2}$
and $m_{\rm eff}^2$ is negative when $\epsilon_0>\frac{3}{2}$.
From now on, we suppose $\lambda_g>-\frac{1}{4}\xi a_g$ so that
$H_0^2>0$ is satisfied.

We know that $g(\epsilon)=0$ has positive roots when $\lambda_f\leq \lambda_+$.
We first consider $\lambda_f=\lambda_+$ case where there exists a multiple positive root $\epsilon_*$. 
Since we supposed $\lambda_g > -\frac{1}{4}\xi a_g$,
we can evaluate $\epsilon_*$ as
\begin{equation}
 \frac{3}{2}-\epsilon_*
=\frac{3 (\frac{\lambda_g}{\xi}+\frac{a_g}{4})}
{(\frac{\lambda_g}{\xi}+\frac{a_g}{4})+\frac{9}{4}a_g+(1-a_g)+\sqrt{\bar{c}^2+9a_g(1-a_g)}}
>0 \ .
\label{e*32}
\end{equation}
Then, $m_{\rm eff}^2$ is positive. Therefore, we find that the de Sitter solution corresponding to the multiple root is stable.

Next, we decrease $\lambda_f$ from $\lambda_+$.
The inner root always satisfies $\epsilon_{\mathrm{in}}<\epsilon_*$ 
as we mentioned in subsection \ref{subs-h02}.
We know $\epsilon_*$ is smaller than $\frac{3}{2}$.
Therefore the inner root is always stable since $\epsilon_{\mathrm{in}}<\epsilon_*<\frac{3}{2}$.
On the other hand,
the outer root always satisfies $\epsilon_{\mathrm{out}}>\epsilon_*$.
Since $\epsilon_*$ is smaller than $\frac{3}{2}$
and the outer root monotonically increase as $\lambda_f$ decreases,
we can expect that $\epsilon_{\mathrm{out}}$ sometime reaches $\frac{3}{2}$.
Once $\epsilon_{\mathrm{out}}$ reaches $\frac{3}{2}$, the effective mass vanishes.
There, $\lambda_f$ is given by
\begin{equation}
  \lambda_{\frac{3}{2}}
=\frac{4}{27}\Bigl[3\Bigl(\lambda_{g}+\frac{\xi
a_{g}}{4}\Bigr)+\xi(1-a_{g})\Bigr]>0 \ ,
\label{l32}
\end{equation}
and Hubble scale reads
\begin{equation}
 H_0^2\Bigl(\frac{3}{2}\Bigr)=\lambda_g+\frac{\xi a_g}{4}>0 \ .
\end{equation}
Note that $\lambda_{p}<\lambda_{\frac{3}{2}}<\lambda_+$ 
because  $\epsilon_*<\frac{3}{2}<\epsilon_{p}$ (see (\ref{ecpm}) and (\ref{e*32})).
Therefore, the outer root is stable when $\lambda_f\geq\lambda_{\frac{3}{2}}$
and unstable when $\lambda_f<\lambda_{\frac{3}{2}}$.

\subsection{Appearance of the Higuchi bound}

In this subsection,
we will evaluate the effective mass of the massive graviton corresponding to the anisotropy.

From the definition of $m_{\rm eff}^2$ and the first line of (\ref{h02}), 
we can deduce the following expression
\begin{eqnarray}
 m_{\rm eff}^2 (\epsilon_0) - 2 H_0^2(\epsilon_0)
&=& - \frac{3\xi}{\epsilon_0} \left[ a_g \epsilon_0^2 + \frac{2}{3}\bar{c} \epsilon_0 -(1-a_g)  \right] \nonumber \\
&=&\frac{3\xi a_g}{\epsilon_0}(\epsilon_* - \epsilon_0)(\epsilon_0- \epsilon_2 ) \ ,
\label{meff-2H2}
\end{eqnarray}
where $\epsilon_*$ is given in (\ref{e*}) and we defined
\begin{equation*}
   \epsilon_2  =   \frac{-\bar{c}-\sqrt{\bar{c}^2+9a_g(1-a_g)}}{3a_g}<0 \ .
\end{equation*}
Since $\epsilon_2$ is negative, 
the sign of $m_{\rm eff}^2-2H_0^2$ depends on that of
($\epsilon_* - \epsilon_0$).
Namely,  $\epsilon_0=\epsilon_*$ is equivalent to $m_{\rm eff}^2 = 2H_0^{2}$,
$ \epsilon_0<\epsilon_* $ leads to $ m_{\rm eff}^2 > 2 H_0^2 $, 
and $ \epsilon_0>\epsilon_*$ leads to $ m_{\rm eff}^2  < 2 H_0^2$.
When $\lambda_f = \lambda_+ $, 
the multiple root $\epsilon_*$  
obviously satisfies $m_{\rm eff}^2 = 2H_0^{2}$.
When $\lambda_f<\lambda_+$, there are two positive roots for $g(\epsilon)=0$.
The inner root always satisfies 
$\epsilon_{\mathrm{in}} < \epsilon_*$
as we mentioned in subsection \ref{subs-h02}.
Hence, the inner root always satisfies $m_{\rm eff}^2 > 2H_0^2$ .
On the other hand, the outer root always satisfies 
$\epsilon_{\mathrm{out}} > \epsilon_*$.
Therefore, the outer root always satisfies $m_{\rm eff}^2 < 2 H_0^2$.

Remarkably, the equation $m_{\rm eff}^2 -2H_0^2=0$ coincides 
with the equation determining the multiple root $\epsilon_*$
(see (\ref{multiple}) and (\ref{meff-2H2})).
That is the reason why the bifurcation point of 
de Sitter solutions is exactly the same as the Higuchi bound.

Note that 
the anisotropy decays more rapidly than Hubble time scale $1/H_0$ for
the inner root 
and it decays more slowly than $1/H_0$ or exponentially grows for the outer root
if we use the analysis of Sec. \ref{linear_pert}.

Finally, we shall see that the ratio of the effective mass to Hubble scale 
monotonically varies along the line that the value of $\epsilon_0$ is constant 
on $\lambda_g$-$\lambda_f$ plane.
We define $\zeta$ as the ratio of the effective mass to Hubble scale,
\begin{equation}
 \zeta=\frac{m_{\rm eff}^2}{H_0^2}
 =\frac{\xi[a_g \epsilon_0+(1-a_g)\frac{1}{\epsilon_0}](3-2\epsilon_0)}
{\lambda_g+\xi a_g (2-\epsilon_0)(\epsilon_0-1)} \ .
\end{equation}
From this expression, 
it is obvious that 
$\partial \zeta /\partial \lambda_g|_{\epsilon_0={\rm const.}}<0$ 
for $\epsilon_0>\frac{3}{2}$ where $\zeta$ is positive, 
and $\partial \zeta /\partial \lambda_g|_{\epsilon_0={\rm const.}}>0$ 
for $\epsilon_0<\frac{3}{2}$ where $\zeta$ is negative.

\begin{figure}[!h]
\centering \includegraphics[height=3.5in]{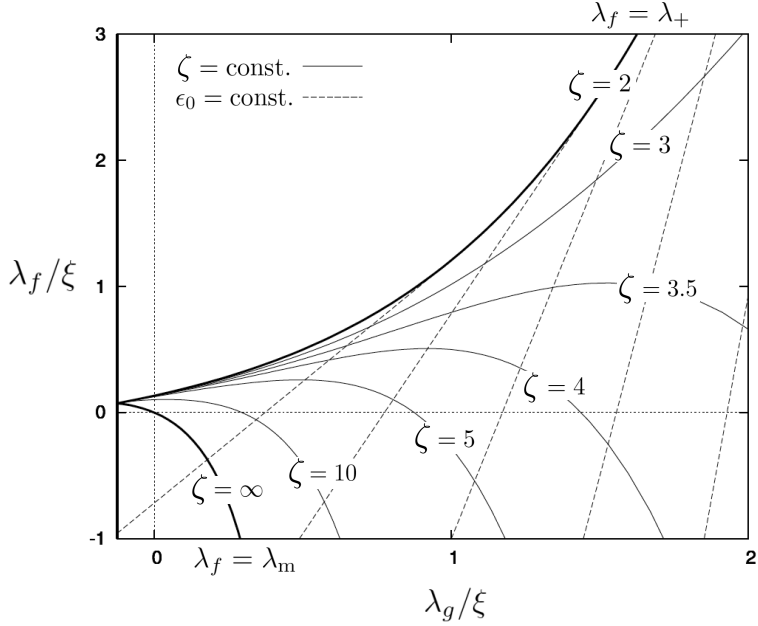}
\caption{We plotted $\zeta=$const. curves of the inner root on $\lambda_g$-$\lambda_f$ plane. 
We set $a_g=0.5$. The inner root is always stable since $m_{\rm eff}^2$ is positive. 
In this figure, $\zeta=2$ and $\zeta=+\infty$ curves coincide with $\lambda_f = \lambda_+$ and $\lambda_f = \lambda_{m}$ curves, respectively.
Note that if we start from a point on $\lambda_f=\lambda_+$,
$\zeta$ monotonically increases along $\epsilon_{\mathrm{in}}=\mathrm{const.}$ line.}
\label{fig.4}
\end{figure}

\begin{figure}[!h]
\centering \includegraphics[height=3.5in]{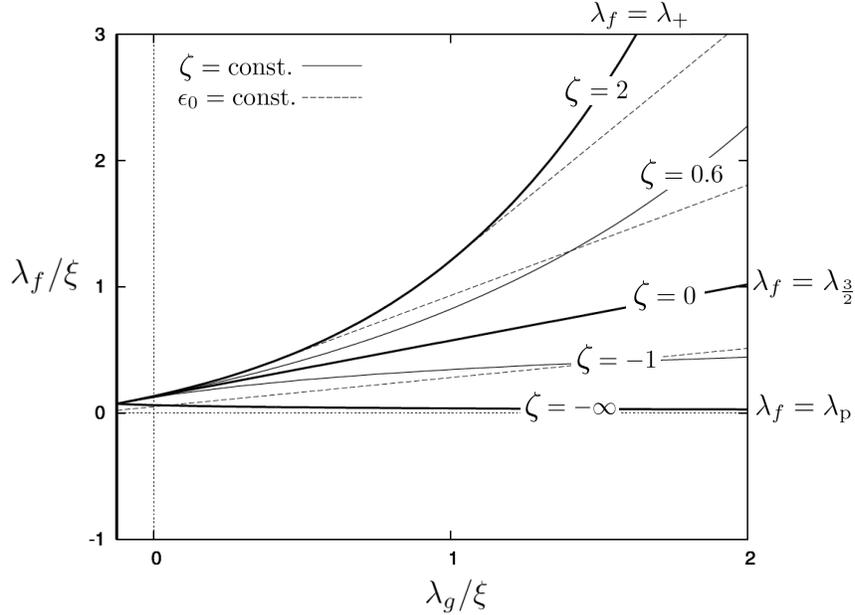}
\caption{We plotted $\zeta=$const. curves of the outer root on $\lambda_g$-$\lambda_f$ plane. 
We set $a_g=0.5$. 
The outer root is stable 
above $\lambda_{\frac{3}{2}}$ where $m_{\rm eff}^2$ is positive
and unstable below $\lambda_{\frac{3}{2}}$ 
where $m_{\rm eff}^2$ is negative. 
In this figure,
$\zeta=2$, $\zeta=0$ and $\zeta=-\infty$ curves coincide with $\lambda_f = \lambda_+$, $\lambda_f = \lambda_{\frac{3}{2}}$
and $\lambda_f = \lambda_{p}$ curves, respectively.
We see that if we start from a point on $\lambda_f=\lambda_+$,
$\zeta$ monotonically decreases along $\epsilon_{\mathrm{out}}=\mathrm{const.}$ line
in the stable region 
and monotonically increases 
in the unstable region. 
}
\label{fig.5}
\end{figure}

We will check
how the line that $\epsilon_0$ is constant can be drawn on $\lambda_g$-$\lambda_f$ plane.
When we fix the value of $\epsilon_0$,
 $g(\epsilon_0)=0$ gives the relation between $\lambda_g$ and $\lambda_f$ as
\begin{equation}
 \lambda_f = \frac{1}{\epsilon_0^2}\lambda_g 
-\xi a_g +\frac{3a_g\xi \epsilon_0^2 + (1-3a_g)\xi \epsilon_0-
\xi(1-a_g)}{\epsilon_0^3} \ .
\end{equation}
On $\lambda_g$-$\lambda_f$ plane,
each point in the region $\lambda_f<\lambda_+$ determines two lines: 
one for the inner root and the other for the outer root.
From the fact that
\begin{equation}
 \frac{\mathrm{d}\lambda_+}{\mathrm{d}\lambda_g}
=\biggl(\frac{\bar{c}+\sqrt{\bar{c}^2+9a_g(1-a_g)}}{3(1-a_g)}\biggr)^2
=\frac{1}{\epsilon_*^2} \ ,
\end{equation}
each line is tangential to $\lambda_f=\lambda_+$ curve.
We also know that $\lambda_f=\lambda_+$ is a convex function since
from
\begin{equation}
 \frac{\mathrm{d}\epsilon_*}{\mathrm{d}\lambda_g}
=\frac{-\epsilon_*}{\xi\sqrt{\bar{c}^2+9a_g(1-a_g)}}<0 \ ,
\end{equation}
we can obtain
\begin{equation}
 \frac{\mathrm{d}^2\lambda_+}{\mathrm{d}\lambda_g^{2}}
=\frac{-2}{\epsilon_*^3} \frac{\mathrm{d}\epsilon_*}{\mathrm{d}\lambda_g}
>0 \ .
\end{equation}
Therefore, the series of lines cover the whole region satisfying $\lambda_f<\lambda_+$.

Using these formulae, we depicted Fig. \ref{fig.4} and Fig. \ref{fig.5}

\section{Conclusion}\label{sec:C}

We investigated the cosmic no-hair conjecture in the presence of spin-2 matter.
More precisely, we studied the cosmic no-hair conjecture in bimetric gravity
 using the perturbative method. 
First, we analyzed de Sitter solutions and found that there are two branches of de Sitter solutions. 
We examined the stability of de Sitter solutions and found that there always at least one stable solution. 
Finally, we evaluated the effective mass and found 
that the stable branch of de Sitter solutions satisfies the Higuchi bound. 
The other branch does not satisfy the Higuchi bound. 
The bifurcation point of two branches exactly coincides with the Higuchi bound. 
Thus, we concluded that there exists a de Sitter solution for which the anisotropy decays 
and the effective mass for these perturbations satisfy the Higuchi bound.
Since the cosmic no-hair conjecture is already violated at the linear
level in known cases, our result indicates that the cosmic no-hair
conjecture is correct in bimetric gravity even though we have not given
the nonlinear analysis.

As a future work,
it would be interesting to explore the meaning behind the curious fact that 
the bifurcation point of two branches of de Sitter solutions coincides with the Higuchi bound. 
Moreover, since we have found that at least one branch of de Sitter solutions is stable in our analysis, 
we can consider inflation in bimetric gravity without pathologies. 
It would be important to clarify what kind of signatures peculiar to bimetric gravity appear 
for example in the cosmic microwave background radiation.

\section*{Acknowledgement}
We would like to thank Emir G$\ddot{\rm u}$mr$\ddot{\rm u}$k\c{c}$\ddot{\rm u}$o$\breve{\rm g}$lu, Kei-ichi Maeda, 
Shinji Mukohyama, Claudia de Rham, Takahiro Tanaka, and Andrew J. Tolley for fruitful discussions.  JS is supported by  the
Grant-in-Aid for  Scientific Research Fund of the Ministry of 
Education, Science and Culture of Japan (C) No.22540274, (A) (No. 21244033, No.22244030), the Grant-in-Aid for  Scientific Research on Innovative Area No.21111006, and
JSPS under the Japan-Russia Research Cooperative Program. TT is supported by
the Japan Society for the Promotion of Science
(JSPS) grant No. 23 - 661. This work is partially supported by  
 the Grant-in-Aid for the Global COE Program 
``The Next Generation of Physics, Spun from Universality and Emergence".

\appendix
\section{Derivation of Basic equations}\label{sec:DB}

In this appendix, we derive basic equations.

\subsection{Ansatz and Lagrangian}

We start with the anisotropic metric ansatz for $g_{\mu\nu}$ and $f_{\mu\nu}$
\begin{equation*}
	ds^{2}
=-N^{2}(t)dt^{2}+e^{2\alpha(t)}[e^{-4\sigma(t)}dx^{2}+e^{2\sigma(t)}(dy^{2}+dz^{2})] \ ,
\end{equation*}
and
\begin{equation*}
	ds'^{2}
=-M^{2}(t)dt^{2}+e^{2\beta(t)}[e^{-4\lambda(t)}dx^{2}+e^{2\lambda(t)}(dy^{2}+dz^{2})]  \ ,
\end{equation*}
respectively.
From these metrics, scalar curvatures are calculated as
\begin{equation}
 R[g_{\mu\nu}] = \frac{1}{N^{2}}(-6\dot{\alpha}^{2}+6\dot{\sigma}^{2}) \ , \qquad
 R[f_{\mu\nu}] = \frac{1}{M^{2}}(-6\dot{\beta}^{2}+6\dot{\lambda}^{2}) \ .
\end{equation}
Moreover, $g^{-1}f$ is given by
\begin{align*}
 g^{-1}f &=
 \begin{pmatrix}
  (M/N)^{2} & & & \\
  & e^{2\beta-2\alpha-4\lambda+4\sigma} & & \\
  & & e^{2\beta-2\alpha+2\lambda-2\sigma} & \\
  & & & e^{2\beta-2\alpha+2\lambda-2\sigma}
 \end{pmatrix}
&=
 \begin{pmatrix}
  \gamma^{2} & & &\\
  & A^{2} & & \\
  & & B^{2} & \\
  & & & B^{2} 
 \end{pmatrix} \ ,
\end{align*}
where we have defined variables as
\begin{equation*}
 \gamma = M/N \ , \qquad
 \epsilon=e^{\beta-\alpha} \ ,\qquad
 \eta=e^{\lambda-\sigma} \ ,
\end{equation*}
\begin{equation*}
 A=\epsilon\eta^{-2}=e^{\beta-\alpha-2\lambda+2\sigma} \ ,\qquad
 B=\epsilon\eta=e^{\beta-\alpha+\lambda-\sigma} \ .
\end{equation*}
Thus, we obtain
\begin{equation}
  L = 1-\sqrt{g^{-1}f}=
 \begin{pmatrix}
  1-\gamma & & &\\
  & 1-A & & \\
  & & 1-B & \\
  & & & 1-B 
 \end{pmatrix} \ ,
\end{equation}
Then, we can calculate the interaction term as
\begin{eqnarray}
 F_{2}&=&\frac{1}{2}\bigl([L]^{2}-[L^{2}]\bigr)\nonumber\\
 &=&\frac{1}{2}[(4-A-2B-\gamma)^{2}-(4-2A-4B+A^{2}+2B^{2}-2\gamma+\gamma^{2})]\nonumber\\
 &=&[6-3A-6B+B(2A+B)+\gamma(-3+A+2B)] \ .
\end{eqnarray}
Therefore, the Lagrangian reads
\begin{eqnarray}
 \mathcal{L}=
&& M_{g}^{2}e^{3\alpha}\Bigl[\frac{3}{N}(-\dot{\alpha}^{2}+\dot{\sigma}^{2})-N\Lambda_{g}\Bigr]
  +M_{f}^{2}e^{3\beta}\Bigl[\frac{3}{M}(-\dot{\beta}^{2}+\dot{\lambda}^{2})-M\Lambda_{f}\Bigr]
\nonumber\\
&& +m^{2}M_{e}^{2}Ne^{3\alpha}\bigl[6-3A-6B+B(2A+B)
 +\gamma(-3+A+2B)\bigr] \ .
\end{eqnarray}

\subsection{Equations of motion and constraints}

We normalize parameters and time with $M_e$ as follows:
\begin{equation}
 a_{g}=\frac{M_{e}^{2}}{M_{g}^{2}} \ ,\qquad
 \xi=\frac{m^{2}}{M_{e}^{2}} \ , \qquad
 \lambda_{g}=\frac{\Lambda_{g}}{3M_{e}^{2}} \ , \qquad
 \lambda_{f}=\frac{\Lambda_{f}}{3M_{e}^{2}} \ , \qquad
 ' = \cdot/M_{e} \ .\nonumber
\end{equation}
Note that $0<a_{g}<1$ from the definition of $M_e$.

From the Lagrangian, we obtain the equations of motion
\begin{equation}
 \Bigl(\frac{\alpha'}{N}\Bigr)'+3\frac{\sigma'^{2}}{N}
+\frac{1}{6}\xi a_{g}[N(3A+6B-2B(2A+B))-M(9-2A-4B)]=0 \ ,
\end{equation}
\begin{equation}
 \Bigl(\frac{\beta'}{M}\Bigr)'+3\frac{\lambda'^{2}}{M}-\frac{1}{6}\xi (1-a_{g})\frac{1}{\epsilon^{3}}[N(3A+6B-2B(2A+B))-M(9-2A-4B)]=0 \ ,
\end{equation}
\begin{equation}
 \Bigl(\frac{\sigma'}{N}\Bigr)'+3\frac{\alpha'\sigma'}{N}
  +\frac{1}{3}\xi a_{g}(A-B)[N(3-B)-M]=0 \ , \label{eom_s}
\end{equation}
\begin{equation}
  \Bigl(\frac{\lambda'}{M}\Bigr)'+3\frac{\beta'\lambda'}{M}
  -\frac{1}{3}\xi
  (1-a_{g})\frac{1}{\epsilon^{3}}(A-B)[N(3-B)-M]=0 \ , \label{eom_l}
\end{equation}
and the constraints
\begin{equation}
 \Bigl(\frac{\alpha'}{N}\Bigr)^{2}-\Bigl(\frac{\sigma'}{N}\Bigr)^{2}=\lambda_{g}+\frac{1}{3}\xi
 a_{g}[-6+3A+6B-B(2A+B)] \ ,
\end{equation}
\begin{equation}
 \Bigl(\frac{\beta'}{M}\Bigr)^{2}-\Bigl(\frac{\lambda'}{M}\Bigr)^{2}
 =\lambda_{f}+\frac{1}{3}\xi(1-a_{g})\frac{1}{\epsilon^{3}}(3-A-2B) \ .
\end{equation}
It is easy to find the consistency relation
\begin{equation}
 \frac{M}{N}
=\frac{\beta'(3A+6B-2B(2A+B))-\lambda'(2A-2B)(3-B)}{\alpha'(9-2A-4B)-\sigma'(2A-2B)} \ .
\end{equation}
From the linear combination of (\ref{eom_s}) and (\ref{eom_l}), 
we can also obtain a conserved quantity
\begin{equation}
 E = \frac{1}{a_{g}}\frac{e^{3\alpha}\sigma'}{N}+\frac{1}{1-a_{g}}\frac{e^{3\beta}\lambda'}{M} \ .
\end{equation}


\end{document}